\documentclass{elsart}
\usepackage{amsfonts}
\usepackage{amsmath}
\renewcommand{\Re}{\mathop{\mathrm{Re}}}
\newcommand{\p}{^{\prime}}
\newcommand{\ci}{\mathrm{i}}
\newcommand{\CauchyIntX}{\makebox{$\displaystyle -\hspace{-0.95em}\int_0^1$}}
\newcommand{\cauchyInt}{f}
\newcommand{\ksih}{\xi_{\scriptscriptstyle H}}
\newcommand{\ksix}{\xi_{\scriptscriptstyle X}}
\newcommand{\ksiy}{\xi_{\scriptscriptstyle Y}}
\newcommand{\etax}{\eta_{\scriptscriptstyle X}}
\newcommand{\etay}{\eta_{\scriptscriptstyle Y}}
\begin{document}
\runauthor{Rutily, Chevallier and Bergeat}
\runtitle{Integral equations for the $H$- $X$- and $Y$-functions}
\begin{frontmatter}
\title{Integral equations satisfied by the $H$- $X$- and $Y$-functions and the albedo problem}
\author{Bernard Rutily\thanksref{br}},
\author{Lo\"{\i}c Chevallier} and
\author{Jacques Bergeat}
\address{Transfer Group, Centre de Recherche Astronomique de Lyon\\ (UMR 5574 du CNRS),
Observatoire de Lyon, 9 avenue Charles Andr\'{e},\\ 69561 Saint-Genis-Laval Cedex, France}
\thanks[br]{Corresponding author. Tel.: 33 4 78 86 83 79\\
{\em Email address\/}: rutily@obs.univ-lyon1.fr (B. Rutily)}
\date{7 March 2003}
\received{7 March 2003}
\accepted{9 July 2003}
\begin{abstract}
We come back to a non linear integral equation satisfied by the function $H$,
which is distinct from the classical $H$-equation. Established for the first time by
Busbridge (1955), it appeared occasionally in the literature since then. First of all, this
equation is generalized over the whole complex plane using the method of residues.
Then its counterpart in a finite slab is derived; it consists in two series of integral
equations for the $X$- and $Y$-functions. These integral equations are finally applied to the
solution of the albedo problem in a slab.
\end{abstract}
\begin{keyword}
Transport (or transfer) theory; plane-parallel geometry; isotropic scattering;
integral equations; functions $H$, $X$ and $Y$; albedo problem
\end{keyword}
\end{frontmatter}

\section{Introduction}
 This paper originates in a work by Busbridge \cite{busbridge1955a} on the calculation of the
classical $X$- and $Y$-functions of transfer theory. A non linear integral equation for the
$H$-function is proved in this article, that is not the classical $H$-equation: see Eqs.
(5.6)-(5.7) in \cite{busbridge1955a}, hereafter Eqs. (\ref{eq27}) and (\ref{eq29}).\\
 In the second of a series of papers submitted for publication \cite{rutily2002,rutily1,rutily2}, we deduced
Busbridge's equation from the properties of the $N$-function, which can be seen as the
"heart" of the $H$-function. The simplifying role of Busbridge's equation for solving the
singular $H$-equation was stressed in Sec. III-A of \cite{rutily1}. It also clarifies the solution of the
albedo problem, yielding the surface field from the internal field of a half-space
illuminated from outside. It thus provides the link between the internal and the surface
solution to the albedo problem in a half-space.\\
 Considering the importance of Busbridge's equation in semi-infinite geometry,
it is tempting to generalize it in the finite case and examine the consequence it has on
the albedo problem in a finite slab: this is the object of the present article. After a
reminder on the classical functions $H$, $X$ and $Y$ (Sec. \ref{sec2}), we come back in Sec. \ref{sec3} to
Busbridge's $H$-equation. This equation is generalized to the complex plane and proved
by the method of residues. This approach is then continued in a finite slab, which leads
to two series of integral equations for the $X$- and $Y$-functions (Sec. \ref{sec4}). An application
to the slab albedo problem is proposed in Sec. \ref{sec5}.

\section{Classical integral equations for the $H$-, $X$- and $Y$-functions}\label{sec2}

 We consider a slab of constant volumic albedo $a \in ]0,1[$ and optical thickness
$b$ ($0<b\leq +\infty$), where light scattering is isotropic and monochromatic. The auxiliary
functions describing the radiation field depend on the $a$-parameter in a half-space ($b =
+ \infty$), and on the $a$- and $b$-parameters in a finite slab. This dependence is always explicitly mentioned.

\subsection{Integral equations for the $H$-function in a half-space}

 We introduce the dispersion function for the adopted scattering law
\begin{equation}
\label{eq1}
T(a,z) = 1 + {\frac{a}{2}} z \int_{-1}^{+1} {\frac{\d u}{u - z}} \qquad (z \neq \pm 1),
\end{equation}
which describes the multiple scattering of photons in an unbounded medium. This
function is defined in the whole complex plane $\mathbb{C}$, except at $\pm 1$, provided that the
integral is calculated in the sense of the Cauchy principal value when $z \in ]-1,+1[$. When
$0<a<1$ the characteristic equation $T(a,z)=0$ has four non zero roots in $\mathbb{C}$, namely
two pairs of opposite real numbers since the $T$-function is even \cite{busbridge1960,case1967}.
They are denoted by $\pm 1/k$ ($0<k<1$) and $\pm 1/K$ ($0<K<1$). The important root $k$ in $]0,1[$ is
calculated by solving the transcendental equation
\begin{equation}
\label{eq2}
T(a,1/k) = 1 - {\frac{a}{2}}{\frac{1}{k}} \ln \frac{1+k}{1-k} = 0 \qquad (0<k<1),
\end{equation}
which has an exact solution $k = k(a)$ \cite{siewert1999}.\\
 The $H$-function may be defined as the unique solution, analytic in the right
complex half-plane, of any of the following integral equations 
\begin{equation}
\label{eq3}
H(a,z) [1 - \ksih(a,z)] = 1 \qquad (z \neq -1),
\end{equation}
\begin{equation}
\label{eq4}
T(a,z) H(a,z) = 1 - \ksih(a,-z) \qquad (z \neq \pm 1), 
\end{equation}
\begin{equation}
\label{eq5}
[1 - \ksih(a,z)] [1 - \ksih(a,-z)] = T(a,z) \qquad (z \neq \pm 1),
\end{equation}
where
\begin{equation}
\label{eq6}
\ksih(a,z) = \frac{a}{2} z \int_{0}^{1} H(a,u) \frac{\d u}{u+z} \qquad (z \neq -1)
\end{equation}
(the integral is a Cauchy principal value for $z \in ]-1,0[$). \\
 Putting $z = +1/k$ and $z = +1/K$ into Eq. (\ref{eq4}) and observing that the $H$-function
does not diverge at these points, one obtains $1 - \ksih (a,-1/k) = 1 - \ksih(a,-1/K) = 0$.
Equation (\ref{eq3}) then shows that $H$ diverges at $z = -1/k$ and $z = -1/K$.
This equation also defines the extension of the $H$-function in $\mathbb{C} \setminus \{-1/k,-1,-1/K\}$,
 and we have for any $z$ in $\mathbb{C} \setminus \{-1/k,-1,-1/K\}$
\begin{equation}
\label{eq7}
\frac{1}{H(a,z)} = 1 - \ksih (a,z) = 1 - \frac{a}{2} z \int_{0}^{1} H(a,u) \frac{\d u}{u+z} .
\end{equation}
 Equation (\ref{eq4}) reads also, for any $z\in \mathbb{C} \setminus \{-1/k,\pm 1,-1/K\}$
\begin{equation}
\label{eq8}
T(a,z) H(a,z) = 1 + \frac{a}{2} z \int_{0}^{1} H(a,u) \frac{\d u}{u-z} \,,
\end{equation}
which allows to extend $H$ by continuity at $z = -1$: $H(a,-1) = 0$. Finally, it follows from
Eqs. (\ref{eq3}) and (\ref{eq5}) that for any $z$ in $\mathbb{C} \setminus \{\pm 1/k,\pm 1,\pm 1/K\}$
\begin{equation}
\label{eq9}
T(a,z) H(a,z) H(a,-z) = 1 .
\end{equation}
 Equations (\ref{eq7})-(\ref{eq9}) are the regular $H$-equation \cite{ambartsumian1942}, the singular $H$-equation \cite{sobolev1949}, and
the factorization relation \cite{crum1947} respectively. We refer here to those papers where, to our
knowledge, each equation can be found for the first time.\\
 The behavior of the $H$-function at infinity is required in Sec. \ref{sec3}. We recall that
\begin{equation}
\label{eq10}
H(a,\infty) = \frac{1}{\sqrt{1-a}} ,
\end{equation}
as can be seen by putting $z \to \infty$ in Eqs. (\ref{eq4}) and (\ref{eq5}), observing that $T(a,\infty) = 1-a$.

\subsection{Integral equations for the functions $X$ and $Y$ in a finite slab}

 The $X$- and $Y$-functions are defined in $\mathbb{C}^* = \mathbb{C} \setminus \{0\}$ and related by
\begin{equation}
\label{eq11}
X(a,b,z) = Y(a,b,-z) \exp(-b/z) ,\; Y(a,b,z) = X(a,b,-z) \exp(-b/z) .
\end{equation}
 They are the unique solution, analytic in the right complex half-plane, of coupled
integral equations involving the auxiliary functions 
\begin{equation}
\label{eq12}
\ksix(a,b,z) = \frac{a}{2} z \int_{0}^{1} X(a,b,u) \frac{\d u}{u+z} \,, 
\end{equation}
\begin{equation}
\label{eq13}
\ksiy (a,b,z) = \frac{a}{2} z \int_{0}^{1} Y(a,b,u) \frac{\d u}{u+z} \,, 
\end{equation}
defined over $\mathbb{C} \setminus \{-1\}$. These equations are either the regular $X$- and $Y$-equations \cite{ambartsumian1943}
\begin{equation}
\label{eq14}
X(a,b,z) [1 - \ksix (a,b,z)] + Y(a,b,z) \ksiy (a,b,z) = 1 \qquad (z \neq 0,-1),
\end{equation}
\begin{equation}
\label{eq15}
X(a,b,z) \ksiy (a,b,-z) + Y(a,b,z) [1 - \ksix (a,b,-z)] = \exp(-b/z) \quad (z \neq 0,+1),
\end{equation}
or the singular $X$- and $Y$-equations \cite{sobolev1949}
\begin{equation}
\label{eq16}
T(a,z) X(a,b,z) = 1 - \ksix (a,b,-z) - \ksiy (a,b,z) \exp(-b/z) \quad (z \neq 0,\pm 1),
\end{equation}
\begin{equation}
\label{eq17}
T(a,z) Y(a,b,z) = [1 - \ksix (a,b,z)] \exp(-b/z) - \ksiy (a,b,-z) \quad (z \neq 0,\pm 1). 
\end{equation}
We have furthermore the relation \cite{sobolev1957}
\begin{equation}
\label{eq18}
[1 - \ksix (a,b,z)] [1 - \ksix (a,b,-z)] - \ksiy (a,b,z) \ksiy (a,b,-z) = T(a,z) \quad (z \neq \pm 1),
\end{equation}
which generalizes the factorization relation (\ref{eq9}) for semi-infinite media.\\
 Let $b$ go to infinity and suppose that $\Re(z)>0$, where $\Re(z)$ is the real part of
the complex number $z$. Then $X(a,b,z) \to H(a,z)$ and $Y(a,b,z) \to 0$, so that Eqs. (\ref{eq14}),
(\ref{eq16}) and (\ref{eq18}) yield Eqs. (\ref{eq3})-(\ref{eq5}) again respectively. Equations (\ref{eq15}) and (\ref{eq17}) reduce to $0
= 0$.\\
 The behavior of the functions $X$ and $Y$ in a neighborhood of 0 and at infinity
follows immediately from Eqs. (\ref{eq16}) and (\ref{eq17}). Let $z$ tends to 0 in these equations. Then
$T(a,z)$ and $1-\ksix(a,b,z)$ tend to 1, and $\ksiy(a,b,z)$ tends to $(a/2)\beta_{\mathrm{-1}}(a,b)z$, where $\beta_{\mathrm{-1}}(a,b)$ is the moment of order -1 of the $Y$-function, viz.
\begin{equation}
\label{eq19}
\beta_{-1} (a,b) = \int_{0}^{1} Y(a,b,u) \frac{\d u}{u} .
\end{equation}
Consequently
\begin{equation}
\label{eq20}
X(a,b,z) \sim 1 - \frac{a}{2} \beta_{-1} (a,b) z \exp(-b/z) \qquad (z \to 0),
\end{equation}
\begin{equation}
\label{eq21}
Y(a,b,z) \sim \exp(-b/z) + \frac{a}{2} \beta_{-1} (a,b) z \qquad (z \to 0).
\end{equation}
These relations show that $X(a,b,z) \to \infty$ like $-(a/2)\beta_{\mathrm{-1}}(a,b)z \exp(-b/z)$ and
$Y(a,b,z) \to \infty$ like $\exp(-b/z)$ when $z \to 0$ in the left complex half-plane [$\Re(z)<0$], and $X(a,b,z) \to 1$ and $Y(a,b,z) \to 0$ when $z \to 0$ in the
right complex half-plane [$\Re(z) > 0$].\\
 Now letting $z \to \infty$ into Eqs. (\ref{eq12})-(\ref{eq13}) yields
\begin{equation}
\label{eq22}
\ksix (a,b,\infty) = \frac{a}{2} \alpha_{0} (a,b) ,\; \ksiy (a,b,\infty) = \frac{a}{2} \beta_{0} (a,b) ,
\end{equation}
where $\alpha_{0}(a,b)$ and $\beta_{0}(a,b)$ are the moments of order 0 of the $X$- and $Y$-functions
\begin{equation}
\label{eq23}
\alpha_{0} (a,b) = \int_{0}^{1} X(a,b,u) \d u ,\; \beta_{0} (a,b) = \int_{0}^{1} Y(a,b,u) \d u .
\end{equation}

\section{Busbridge's integral equation for the $H$-function}\label{sec3}

This integral equation was established in 1955 by Busbridge \cite{busbridge1955a} as an auxiliary
result for the calculation of the $X$- and $Y$-equations. This first appearance was so discreet
that Busbridge didn't include her lemma in \cite{busbridge1960}. Busbridge's equation was "rediscovered"
by Andreasian and Danielian \cite{andreasian1978}, and Das \cite{das1979} wrote a version which needs
simplification. It appears also in the Appendix D of \cite{ivanov1994}. The domain of this equation
has been extended to the real axis in \cite{bergeat1991} and to the complex plane in \cite{rutily1992} and \cite{rutily1}.
In the two references mentioned at last, it is noted that Busbridge's equation clarifies the
albedo problem, since it allows deriving the surface source function from the internal
one, a result we intend to generalize in a finite slab (Sec. \ref{sec5}).\\
 Here we summarize the method used in \cite{bergeat1991} and \cite{rutily1992}, which is based on the
residue theorem. It is well-suited to the main aim of the present paper -- the
generalization of Busbridge's equation in a finite slab -- since it still holds in the finite
case. It consists in integrating the function
\begin{displaymath}
\label{eqn1}
f : z\p \in \mathbb{C} \to H(a,1/z\p) \frac{1}{1-zz\p}
\end{displaymath}
along the contour of Fig. A1 in \cite{bergeat1991}, replacing $H(a,+1/z\p)$ in the left complex half-plane by $1/[T(a,+1/z\p) H(a,-1/z\p)]$.
 Results involve the function
\begin{equation}
\label{eq24}
g(a,v) = \frac{1}{T^2 (a,v) + (\pi av/2)^2} \qquad (0 \leq v < 1),
\end{equation}
which arises when skirting the $]-\infty,-1[$ cut, and the coefficients
\begin{equation}
\label{eq25}
R(a,k) = \frac{1 - k^2}{k^2 + a - 1} ,
\end{equation}
\begin{equation}
\label{eq26}
S(a,k) = 1 - \frac{a k^2}{(k^2 + a - 1)(1 - k^2)} ,
\end{equation}
related to the pole $z = - k$ within the contour. We recall that $k = k(a)$ is the unique
solution in $]0,1[$ of the transcendental Eq. (\ref{eq2}).\\
 It follows from the residue theorem that, for $z \in \mathbb{C} \setminus \{ [-1,0[\cup\{-1/k\} \}$,
\begin{equation}
\label{eq27}
H(a,z) = 1 + \frac{R(a,k)}{H(a,+1/k)} \frac{k z}{1 + k z} + \frac{a}{2} z \int_{0}^{1} (g/H)(a,v) \frac{\d v}{v+z} \,,
\end{equation}
where $(g/H)(a,v) = g(a,v)/H(a,v)$ is a notation we adopt hereafter.\\
This equation is still valid at $z = -1$, provided that $H(a,z)$ is replaced by its
extension by continuity at $z = -1$, which is 0. Hence the exact value of the integral
\begin{equation}
\label{eq28}
\frac{a}{2} \int_{0}^{1} (g/H)(a,v) \frac{\d v}{v - 1} = 1 - \frac{R(a,k)}{H(a,+1/k)} \frac{k}{1-k} .
\end{equation}
For $z = u \in ]-1,0[$, the left-hand side of Eq. (\ref{eq27}) must be replaced by $(gT/H)(a,-u)$ because of the presence of the pole $1/u$ on the cut $]-\infty,-1[$.
Moreover, the integral on the right-hand side of Eq. (\ref{eq27}) has to be calculated in the sense of the Cauchy principal value (symbol $\cauchyInt$).
Hence
\begin{equation}
\label{eq29}
(gT/H)(a,-u) = 1 + \frac{R(a,k)}{H(a,+1/k)} \frac{ku}{1+ku} +\frac{a}{2} u \CauchyIntX (g/H) (a,v) \frac{\d v}{v+u} \,,
\end{equation}
which yields Eq. (\ref{eq28}) again as $u \to -1$ on the right.\\
Finally, Eq. (\ref{eq27}) can be extended by continuity at $z = -1/k$ by
\begin{equation}
\label{eq30}
\frac{a}{2} \int_{0}^{1} (g/H)(a,v) \frac{\d v}{1-kv} = - 1 - \frac{R(a,k)}{H(a,+1/k)}  \left[ S(a,k) + \frac{1}{k} \frac{H\p(a,+1/k)}{H(a,+1/k)} \right] ,
\end{equation}
where $H\p(a,+1/k)$ is the derivative of the $H$-function at $+1/k$.
It can be deduced from the derivative of the $H$-equation (\ref{eq7}) with respect to $z$.\\
Letting $z \to \infty$ into Eq. (\ref{eq27}) and using Eq. (\ref{eq10}), we obtain
\begin{equation}
\label{eq31}
\frac{a}{2} \int_{0}^{1} (g/H)(a,v) \d v = \frac{1}{\sqrt{1-a}} - 1 - \frac{R(a,k)}{H(a,+1/k)} .
\end{equation}
 A new series of integral equations can be deduced from the previous one by
substracting the above equation to Eqs. (\ref{eq27})-(\ref{eq30}): see Sec. III-B of \cite{rutily1}. \\
 The equations demonstrated by Busbridge \cite{busbridge1955a} are (\ref{eq27}) and (\ref{eq29}).
Equation (\ref{eq27}) is also in Refs. \cite{andreasian1978,das1979,ivanov1994,bergeat1991,rutily1992}.
Equation (\ref{eq29}) was retrieved in \cite{bergeat1991} and \cite{rutily1992}, which contain also
Eqs. (\ref{eq28}), (\ref{eq30}) and (\ref{eq31}). The whole set of equations and their immediate consequences
is derived in \cite{rutily1} from a preliminary study of the $N$-function \cite{rutily2002}. \\
 Our focus now is to write the counterparts of Eqs. (\ref{eq27})-(\ref{eq31}) in a finite slab (Sec.
\ref{sec4}), then to investigate some application for the solution of the albedo problem (Sec. \ref{sec5}).

\section{The finite case: integral equations for the $X$- and $Y$-functions}\label{sec4}

 These equations can be proved by the method of residues, using the same contour
as in the semi-infinite case (Fig. A1 in \cite{bergeat1991}). Just integrate the functions
\begin{displaymath}
\label{eqn2}
z\p \in \mathbb{C} \to X(a,b,1/z\p) \frac{1}{1 - zz\p}
\end{displaymath}
\begin{displaymath}
\label{eqn3}
z\p \in \mathbb{C} \to Y(a,b,1/z\p) \frac{1}{1 - zz\p}
\end{displaymath}
along this contour, writing in the left complex half-plane
\begin{multline}
\label{eq32}
\int X(a,b,1/z\p) \frac{\d z\p}{1 - zz\p} = \int \frac{1}{T(a,1/z\p)}  \left\{ [1 - \ksix (a,b,-1/z\p)] \frac{1}{1 - zz\p} \phantom{\frac{\exp(b z\p)}{1 + zz\p}} \right.  \\
\left. - \ksiy (a,b,-1/z\p) \frac{\exp(b z\p)}{1 + zz\p}  \right\} \d z\p ,
\end{multline}
\begin{multline}
\label{eq33}
\int Y(a,b,1/z\p) \frac{\d z\p}{1 - zz\p} = \int \frac{1}{T(a,1/z\p)}  \left\{ [1 - \ksix (a,b,-1/z\p)] \frac{\exp(b z\p)}{1 + zz\p} \right.  \\
\left. \phantom{\frac{\exp(b z\p)}{1 + zz\p}} - \ksiy (a,b,-1/z\p) \frac{1}{1 - zz\p} \right\} \d z\p ,
\end{multline}
due to Eqs. (\ref{eq16})-(\ref{eq17}). This transformation overcomes the difficulty arising from the
divergence of the functions $z\p \to X(a,b,1/z\p)$ and $z\p \to Y(a,b,1/z\p)$ when $z\p \to \infty$ in the left
complex half-plane [see Eqs. (\ref{eq20})-(\ref{eq21})]. One obtains, for $z \in \mathbb{C} \setminus \{[-1,+1] \cup \{\pm 1/k\} \}$
\begin{multline}
\label{eq34}
X(a,b,z) + \frac{1}{T(a,z)} \ksiy (a,b,z) \exp(-b/z) = \frac{1 - \ksix (a,b,-z)}{T(a,z)}  \\
= 1 + R(a,k)  \left\{ [1 - \ksix (a,b,+1/k)] \frac{kz}{1+kz} - \ksiy (a,b,+1/k) \exp(-kb) \frac{kz}{1-kz}  \right\} \\
+ \frac{a}{2} z \int_{0}^{1} g(a,v)  \left\{ [1 - \ksix (a,b,v)] \frac{1}{v+z} - \ksiy (a,b,v) \frac{\exp(-b/v)}{v-z}  \right\} \d v ,
\end{multline}
\begin{multline}
\label{eq35}
Y(a,b,z) - \frac{1}{T(a,z)} [1 - \ksix (a,b,z)] \exp(-b/z) = - \frac{\ksiy (a,b,-z)}{T(a,z)}  \\
= R(a,k) \left\{ [1 - \ksix (a,b,+1/k)] \exp(-kb) \frac{kz}{1-kz} - \ksiy (a,b,+1/k) \frac{kz}{1+kz} \right\} \\
+ \frac{a}{2} z \int_{0}^{1} g(a,v) \left\{ [1 - \ksix (a,b,v)] \frac{\exp(-b/v)}{v-z} - \ksiy (a,b,v) \frac{1}{v+z} \right\} \d v .
\end{multline}
 At $z = \pm 1$, Eqs. (\ref{eq34})-(\ref{eq35}) still hold, provided that $1/T(a,z)$ is replaced by 0 in
both members of the first equality. This simplification has to be done in the left-hand
side for $z = +1$, and in the right-hand side for $z = -1$. One has, at $z = +1$
\begin{multline}
\label{eq36}
X(a,b,+1) = 1 \\
+ R(a,k) \left\{ [1 - \ksix (a,b,+1/ k)] \frac{k}{1+k} - \ksiy (a,b,+1/k) \exp(-kb) \frac{k}{1-k} \right\} \\
+ \frac{a}{2} \int_{0}^{1} g(a,v) \left\{ [1 - \ksix (a,b,v)] \frac{1}{v+1} - \ksiy (a,b,v) \frac{\exp(-b/v)}{v-1}  \right\} \d v ,
\end{multline}
\begin{multline}
\label{eq37}
Y(a,b,+1) = R(a,k) \left\{ [1 - \ksix (a,b,+1/k)] \exp(-kb) \frac{k}{1-k} \right. \\
\left. - \ksiy (a,b,+1/k) \frac{k}{1+k} \right\} \\
+ \frac{a}{2} \int_{0}^{1} g(a,v) \left\{ [1 - \ksix (a,b,v)] \frac{\exp(-b/v)}{v-1} - \ksiy (a,b,v) \frac{1}{v+1} \right\} \d v ,
\end{multline}
and at $z = -1$
\begin{multline}
\label{eq38}
0 = 1 - R(a,k) \left\{ [1 - \ksix (a,b,+1/k)] \frac{k}{1-k} - \ksiy (a,b,+1/k) \exp(-kb) \frac{k}{1+k} \right\} \\
- \frac{a}{2} \int_{0}^{1} g(a,v) \left\{ [1 - \ksix (a,b,v)] \frac{1}{v-1} - \ksiy (a,b,v) \frac{\exp(-b/v)}{v+1} \right\} \d v ,
\end{multline}
\begin{multline}
\label{eq39}
0 = R(a,k) \left\{ [1 - \ksix (a,b,+1/k)] \exp(-kb) \frac{k}{1+k} - \ksiy (a,b,+1/k) \frac{k}{1-k} \right\} \\
+ \frac{a}{2} \int_{0}^{1} g(a,v) \left\{ [1 - \ksix (a,b,v)] \frac{\exp(-b/v)}{v+1} - \ksiy (a,b,v) \frac{1}{v-1} \right\} \d v .
\end{multline}
 If $z = u \in ]-1,0[ \cup ]0,+1[$, Eqs. (\ref{eq34})-(\ref{eq35}) must be replaced by
\begin{multline}
\label{eq40}
\mathrm{Y}(u) X(a,b,u) = 1 \\
+ R(a,k) \left\{ [1 - \ksix (a,b,+1/k)] \frac{ku}{1+ku} - \ksiy (a,b,+1/k) \exp(-kb) \frac{ku}{1-ku} \right\} \\
- (gT)(a,u) \left\{ \mathrm{Y}(-u) [1 - \ksix (a,b,-u)] + \mathrm{Y}(u) \ksiy (a,b,u) \exp(-b/u) \right\} \\
+ \frac{a}{2} u \CauchyIntX g(a,v) \left\{ [1 - \ksix (a,b,v)] \frac{1}{v+u} - \ksiy (a,b,v) \frac{\exp(-b/v)}{v-u} \right\} \d v ,
\end{multline}
\begin{multline}
\label{eq41}
\mathrm{Y}(u) Y(a,b,u) = \\
R(a,k) \left\{ [1 - \ksix (a,b,+1/k)] \exp(-kb) \frac{ku}{1-ku} - \ksiy (a,b,+1/k) \frac{ku}{1+ku} \right\} \\
+ (gT)(a,u) \left\{ \mathrm{Y}(u) [1 - \ksix (a,b,u)] \exp(-b/u) + \mathrm{Y}(-u) \ksiy (a,b,-u) \right\} \\
+ \frac{a}{2} u \CauchyIntX g(a,v) \left\{ [1 - \ksix (a,b,v)] \frac{\exp(-b/v)}{v-u} - \ksiy (a,b,v) \frac{1}{v+u} \right\} \d v ,
\end{multline}
where Y denotes the Heaviside function: $\mathrm{Y}(u)=0 \; \mathrm{for} \; u<0$ and $\mathrm{Y}(u)=1\;\mathrm{for}\;u>0$. These expressions can be extended by continuity at $u = \pm 0$: we find $X(a,b,+0) = 1$ and $Y(a,b,+0) = 0$ when $u \to 0$ on the right, and $0 = 0$ when $u \to 0$ on the left. Moreover, Eqs. (\ref{eq40})-(\ref{eq41}) yield once again Eqs. (\ref{eq36})-(\ref{eq39}) at $u = \pm 1$, provided that $(gT)(a,u)$ is replaced by 0 and the Cauchy integrals are replaced by ordinary integrals.\\
 Finally, putting $z \to -1/k$ into Eqs. (\ref{eq34})-(\ref{eq35}) leads to 
\begin{multline}
\label{eq42}
X(a,b,-1/k) = 1 + R(a,k) \left\{ [S(a,k) + kb] [1 - \ksix (a,b,+1/k)] \phantom{\frac{1}{2}} \right. \\
\left. + \frac{1}{2} \ksiy (a,b,+1/k) \exp(-kb) + \frac{1}{k} \ksiy\p (a,b,-1/k) \exp(+kb) \right\} \\
+ \frac{a}{2} \int_{0}^{1} g(a,v) \left\{ [1 - \ksix (a,b,v)] \frac{1}{1-kv} + \ksiy (a,b,v) \frac{\exp(-b/v)}{1+kv} \right\} \d v ,
\end{multline}
\begin{multline}
\label{eq43}
Y(a,b,-1/k) = - R(a,k) \left\{ \frac{1}{2} [1 - \ksix (a,b,+1/k)] \exp(-kb) \right. \\
\left. - \frac{1}{k} \ksix\p (a,b,-1/k) \exp(kb) + [S(a,k) + kb] \ksiy (a,b,+1/k) \right\} \\
- \frac{a}{2} \int_{0}^{1} g(a,v) \left\{ [1 - \ksix (a,b,v)] \frac{\exp(-b/v)}{1+kv} + \ksiy (a,b,v) \frac{1}{1-kv} \right\} \d v ,
\end{multline}
and putting $z$ $\to$ +1/{\em k\/} 
\begin{multline}
\label{eq44}
X(a,b,+1/k) = 1 + R(a,k) \left\{ \frac{1}{2} [1 - \ksix (a,b,+1/k)] \right. \\
\left. + [S(a,k) - kb] \ksiy (a,b,+1/k) \exp(-kb) - \frac{1}{k} \ksiy\p (a,b,+1/k) \exp(-kb) \right\} \\
+ \frac{a}{2} \int_{0}^{1} g(a,v) \left\{ [1 - \ksix (a,b,v)] \frac{1}{1+kv} + \ksiy (a,b,v) \frac{\exp(-b/v)}{1-kv} \right\} \d v ,
\end{multline}
\begin{multline}
\label{eq45}
Y(a,b,+1/k) = - R(a,k) \left\{ [S(a,k) - kb] [1 - \ksix (a,b,+1/k)] \exp(-kb) \phantom{\frac{1}{2}} \right. \\
\left. + \frac{1}{k} \ksix\p (a,b,+1/k) \exp(-kb) + \frac{1}{2} \ksiy (a,b,+1/k) \right\} \\
- \frac{a}{2} \int_{0}^{1} g(a,v) \left\{ [1 - \ksix (a,b,v)] \frac{\exp(-b/v)}{1-kv} + \ksiy (a,b,v) \frac{1}{1+kv} \right\} \d v .
\end{multline}
  It is possible to eliminate the derivatives at $\pm 1/k$ of the functions $\ksix$ and $\ksiy$
which appear in these four equations. Just use the derivative, at $\pm 1/k$, of Eqs. (\ref{eq16})-(\ref{eq17}),
remarking that $T(a,\pm 1/k) = 0$, $T\p(a,\pm 1/k) = \pm k/R(a,k)$, $\ksiy(a,b,-1/k) \exp(kb) = 1 - \ksix(a,b,+1/k)$
and $[1 - \ksix(a,b,-1/k)] \exp(kb) = \ksiy(a,b,+1/k)$.
The latter two identities follow from Eqs. (\ref{eq16})-(\ref{eq17}) at $-1/k$. Therefore, at $z = -1/k$ 
\begin{multline}
\label{eq46}
X(a,b,-1/k) - (1/k) R(a,k) \ksiy\p (a,b,-1/k) \exp(kb) = \\
R(a,k) \left\{ kb [1 - \ksix (a,b,+1/k)] - (1/k) \ksix\p (a,b,+1/k) \right\} ,
\end{multline}
\begin{multline}
\label{eq47}
Y(a,b,-1/k) - (1/k) R(a,k) \ksix\p (a,b,-1/k) \exp(kb) = \\
- R(a,k) [kb \,\ksiy (a,b,+1/k)] + (1/k) \ksiy\p (a,b,+1/k)] ,
\end{multline}
and at $z = +1/k$
\begin{multline}
\label{eq48}
X(a,b,+1/k) + (1/k) R(a,k) \ksiy\p (a,b,+1/k) \exp(-kb) = \\
R(a,k) [- \ksiy (a,b,+1/k) kb \exp(-kb) + (1/k) \ksix\p (a,b,-1/k)] ,
\end{multline}
\begin{multline}
\label{eq49}
Y(a,b,+1/k) + (1/k) R(a,k) \ksix\p (a,b,+1/k) \exp(-kb) = \\
R(a,k) \left\{ [1 - \ksix (a,b,+1/k)] kb \exp(-kb) + (1/k) \ksiy\p (a,b,-1/ k) \right\} .
\end{multline}
 Inserting these equations into Eqs. (\ref{eq42})-(\ref{eq45}) leads to
\begin{multline}
\label{eq50}
0 = 1 + R(a,k) \left\{ S(a,k) [1 - \ksix (a,b,+1/k)] + \frac{1}{k} \ksix\p (a,b,+1/k) \right. \\
\left.  + \frac{1}{2} \ksiy (a,b,+1/k) \exp(-kb) \right\} \\
+ \frac{a}{2} \int_{0}^{1} g(a,v) \left\{ [1 - \ksix (a,b,v)] \frac{1}{1-kv} + \ksiy (a,b,v) \frac{\exp (-b/v)}{1+kv} \right\} \d v ,
\end{multline}
\begin{multline}
\label{eq51}
0 = R(a,k) \left\{ \frac{1}{2} [1 - \ksix (a,b,+1/k)] \exp(-kb) \right. \\
\left. + S(a,k) \ksiy (a,b,+1/k) - \frac{1}{k} \ksiy\p (a,b,+1/k) \right\} \\
+ \frac{a}{2} \int_{0}^{1} g(a,v) \left\{ [1 - \ksix (a,b,v)] \frac{\exp(-b/v)}{1+kv} + \ksiy (a,b,v) \frac{1}{1-kv} \right\} \d v ,
\end{multline}
\begin{multline}
\label{eq52}
0 = 1 + R(a,k) \left\{ \frac{1}{2} [1 - \ksix (a,b,+1/k)] - \frac{1}{k} \ksix\p (a,b,-1/k) \right. \\
\left. \phantom{\frac{1}{2}} + S(a,k) \ksiy (a,b,+1/k) \exp(-kb) \right\} \\
+ \frac{a}{2} \int_{0}^{1} g(a,v) \left\{ [1 - \ksix (a,b,v)] \frac{1}{1+kv} + \ksiy (a,b,v) \frac{\exp(-b/v)}{1-kv} \right\} \d v ,
\end{multline}
\begin{multline}
\label{eq53}
0 = R(a,k) \left\{ S(a,k) [1 - \ksix (a,b,+1/k)] \exp(-kb) \phantom{\frac{1}{2}} \right. \\
\left. + \frac{1}{2} \ksiy (a,b,+1/k) + \frac{1}{k} \ksiy\p (a,b,-1/k) \right\} \\
+ \frac{a}{2} \int_{0}^{1} g(a,v) \left\{ [1 - \ksix (a,b,v)] \frac{\exp(-b/v)}{1-kv} + \ksiy (a,b,v) \frac{1}{1+kv} \right\} \d v .
\end{multline}
 It is now possible to eliminate the derivatives appearing into Eqs. (\ref{eq42})-(\ref{eq45}) by
calculating them with the help of Eqs. (\ref{eq50})-(\ref{eq53}), which gives
\begin{multline}
\label{eq54}
X(a,b,-1/k) = 1 + kb R(a,k) \left[ 1 - \ksix (a,b,+1/k) - \ksiy (a,b,+1/k) \frac{\sinh(kb)}{kb} \right] \\
+ \frac{a}{2} \int_{0}^{1} g(a,v) \left\{ [1 - \ksix (a,b,v)] \frac{1-\exp[-(1-kv)b/v]}{1-kv} \right. \\
\left. + \ksiy (a,b,v) \exp(-b/v) \frac{1-\exp [(1+kv)b/v]}{1+kv} \right\} \d v ,
\end{multline}
\begin{multline}
\label{eq55}
Y(a,b,-1/k) = \exp(kb) \\
+ kb R(a,k) \left\{ \left[ 1 - \ksix (a,b,+1/k) \right] \frac{\sinh(kb)}{kb} - \ksiy (a,b,+1/k) \right\} \\
- \frac{a}{2} \int_{0}^{1} g(a,v) \left\{ [1 - \ksix (a,b,v)] \exp(-b/v) \frac{1-\exp[(1+kv)b/v]}{1+kv} \right. \\
\left. + \ksiy (a,b,v)  \frac{1-\exp [-(1-kv)b/v]}{1-kv} \right\} \d v ,
\end{multline}
\begin{multline}
\label{eq56}
X(a,b,+1/k) = 1 \\
+ kb R(a,k) \exp(-kb) \left\{ \left[ 1 - \ksix (a,b,+1/k) \right] \frac{\sinh(kb)}{kb} - \ksiy (a,b,+1/k) \right\} \\
+ \frac{a}{2} \int_{0}^{1} g(a,v) \left\{ [1 - \ksix (a,b,v)] \frac{1-\exp[-(1+kv)b/v]}{1+kv} \right. \\
\left. + \ksiy (a,b,v) \exp(-b/v) \frac{1-\exp [-(1-kv)b/v]}{1-kv} \right\} \d v ,
\end{multline}
\begin{multline}
\label{eq57}
Y(a,b,+1/k) = \exp(-kb) \\
+ kb R(a,k) \exp(-kb) \left[ 1 - \ksix (a,b,+1/k) - \ksiy (a,b,+1/k) \frac{\sinh(kb)}{kb} \right] \\
- \frac{a}{2} \int_{0}^{1} g(a,v) \left\{ [1 - \ksix (a,b,v)] \exp(-b/v) \frac{1-\exp[(1-kv)b/v]}{1-kv} \right. \\
\left. + \ksiy (a,b,v) \frac{1-\exp [-(1+kv)b/v]}{1+kv} \right\} \d v .
\end{multline}
 These expressions do satisfy relation (\ref{eq11}), as it should.\\
 Some additional equations involving the functions $\ksix$ and $\ksiy$ can be derived by
letting $z \to \infty$ in the second equalities (\ref{eq34})-(\ref{eq35}), using Eq. (\ref{eq22}) and observing that $T(a,\infty) = 1-a$.
These are
\begin{multline}
\label{eq58}
\frac{1-(a/2)\alpha_{0}(a,b)}{1-a} = \! 1 + R(a,k) \!\left[ 1 - \ksix (a,b,+1/k) + \ksiy (a,b,+1/k) \exp(-kb) \right] \\
+ \frac{a}{2} \int_{0}^{1} g(a,v) \left[ 1 - \ksix (a,b,v) + \ksiy (a,b,v) \exp(-b/v) \right] \d v ,
\end{multline}
\begin{multline}
\label{eq59}
\frac{(a/2) \beta_{0} (a,b)}{1-a} = R(a,k) \left\{ [1 - \ksix (a,b,+1/k)] \exp(-kb) + \ksiy (a,b,+1/k) \right\} \\
+ \frac{a}{2} \int_{0}^{1} g(a,v) \left\{ [1 - \ksix (a,b,v)] \exp(-b/v) + \ksiy (a,b,v) \right\} \d v .
\end{multline}
 Subtracting these equations from Eqs. (\ref{eq34})-(\ref{eq35}) and their extension at $\pm 1$ and
$\pm 1/k$ leads to a new series of integral equations for the functions $X$ and $Y$ (not written
here).\\
 Equations in this series generalize the ones for the function $H$ in a half-space,
and Eqs. (\ref{eq27})-(\ref{eq31}) are retrieved by letting $b \to +\infty$ into Eqs. (\ref{eq34}), (\ref{eq38}), (\ref{eq40}), (\ref{eq50}) and
(\ref{eq58}) respectively.
Note that the $H$-function is nowhere involved in these equations.
The auxiliary functions of finite spaces can thus be calculated without first deriving those for semi-infinite spaces.\\
 Equations (\ref{eq40})-(\ref{eq41}) are the simplified version of two equations derived by
Danielian \cite{danielan1983} in application to Rogovtsov and Samson's calculation of the resolvent
function in a finite slab \cite{rogovtsov1976}.
These equations are, for $u>0$, {Eq.} (\ref{eq40})$[u]_{u>0} + \exp(-b/u)$ $\times$ Eq. (\ref{eq41})$[-u]_{u>0}$
and Eq. (\ref{eq41})$[u]_{u>0} + \exp(-b/u)$ $\times$ Eq. (\ref{eq40})$[-u]_{u>0}$, where Eq. (\ref{eq40})$[u]_{u>0}$ denotes Eq. (\ref{eq40}) with $u>0$,
and Eq. (\ref{eq40})$[-u]_{u>0}$ denotes Eq. (\ref{eq40}) with $-u$ in place of $u$,
supposing then $u>0$ [ditto with Eq. (\ref{eq41})].
As a result, Danielian's equations contain two times too many terms, a remark that still applies to its expressions of the functions $f$ and
$\tilde{f}$ introduced for solving the transfer equation within a slab: see the end of the next section.

\section{Application to the albedo problem}\label{sec5}

 In this section, we consider the "restricted" albedo problem, which consists in
finding the source function of the diffuse field within a slab illuminated from outside.
The calculation of the specific intensity is left aside. It is supposed that parallel rays are
incident on the top surface $\tau = 0$ at an angle $\arccos u_0$ to the inward normal ($0 < u_0 \leq 1$),
 with unit flux per unit area normal to the rays. Since the mean intensity of the direct
field at level $\tau$ is $(1/4\pi) \exp(-\tau/u_0$), the source function for the diffuse field is given by
\begin{equation}
\label{eq60}
S(a,b,\tau,u_0) = \frac{a}{4\pi} B(a,b,\tau,u_0) ,
\end{equation}
where the function $B = B(a,b,\tau,z)$ is solution to the following integral equation
\begin{equation}
\label{eq61}
B(a,b,\tau,z) = \exp(-\tau/z) + \frac{a}{2} \int_{0}^{b} E_{1} (|\tau-t|) B(a,b,t,z) \d t ,
\end{equation}
with parameter $z \in \mathbb{C}^*$. $E_1$ is the first exponential integral function as defined by 
$E_1(\tau) = \int_{0}^{1} \exp(-\tau/u) (\d u/u) \quad (\tau > 0)$.  \\
 There are many analytical methods for solving the above equation. The oldest
one, which is possibly the most direct, originates in Ambartsumian's solution to the
problem of diffuse reflection of light in a half-space \cite{ambartsumian1942}. His approach was generalized
by Busbridge to a finite slab \cite{busbridge1955b}. A modern version of this method, which applies to
complex values of the $z$-parameter, is described in Sec. 3 of \cite{rutily1994}. Ambartsumian's
pioneer paper contains the first expression of the $B$-function on the boundary layer of
a half-space, viz.
\begin{equation}
\label{eq62}
B(a,0,u_0) = H(a,u_0) .
\end{equation}
 In the finite case, it seems that the surface values of the $B$-function have been
given for the first time by Busbridge, who found in \cite{busbridge1955b}
\begin{equation}
\label{eq63}
B(a,b,0,u_0) = X(a,b,u_0) ,\; B(a,b,b,u_0) = Y(a,b,u_0) .
\end{equation}
 The $B$-function may be calculated within the slab using the following relations,
valid for any $z \in \mathbb{C}^*$ as seen in \cite{rutily1994}. In a semi-infinite medium, Eq. (\ref{eq61}) has a solution
as long as $z \in \mathbb{C}^*$ satisfies the condition $\Re(1/z) > - k$. This solution is 
\begin{equation}
\label{eq64}
B(a,\tau,z) = \mathrm{Y}[-\Re(z)] \frac{\exp(-\tau/z)}{T(a,z)} + H(a,z) z \eta(a,\tau,z) ,
\end{equation}
where the Heaviside function Y is extended by 1/2 at 0: $\mathrm{Y}(0) = 1/2$.
The $\eta$-function is defined for $\tau > 0$ and $z \in \mathbb{C}^*$ by
\begin{equation}
\label{eq65}
\eta(a,\tau,z) = \frac{1}{2\ci\pi} \int_{-\ci\infty}^{+\ci\infty} H(a,1/z\p) \exp(\tau z\p) \frac{\d z\p}{1+z\p z} \,,
\end{equation}
and it is extended by continuity at $\tau = 0$ \cite{rutily1994}.\\
 In a finite slab, one has for $z \in \mathbb{C}^*$
\begin{equation}
\label{eq66}
B(a,b,\tau,z) = X(a,b,z) z \etax (a,b,\tau,z) - Y(a,b,z) z \etay (a, b,\tau,z) ,
\end{equation}
where
\begin{equation}
\label{eq67}
\etax (a,b,\tau,z) = \frac{1}{2\ci\pi} \int_{-\ci\infty}^{+\ci\infty} X(a,b,1/z\p) \exp(\tau z\p) \frac{\d z\p}{1+z\p z} \,,
\end{equation}
\begin{equation}
\label{eq68}
\etay (a,b,\tau,z) = \frac{1}{2\ci\pi} \int_{-\ci\infty}^{+\ci\infty} Y(a,b,1/z\p) \exp(\tau z\p) \frac{\d z\p}{1+z\p z} \,,
\end{equation}
for any $\tau \in ]0,b[$ and $z \in \mathbb{C}^*$. These functions can be extended by continuity at $\tau = 0$
and $\tau = b$ \cite{rutily1994}.\\
 Integrals in the right-hand side of Eqs. (\ref{eq65}), (\ref{eq67}) and (\ref{eq68}) are calculated along
the imaginary axis in the sense of the principal value at infinity, i.e.  $\int_{-\ci\infty}^{+\ci\infty} = \lim_{X \to + \infty} \int_{-\ci X}^{+\ci X}$. 
These are also Cauchy principal values when $z$ is on the imaginary axis.
All integrals can be transformed by the method of residues for the numerical evaluation of the functions $\eta$,
$\etax$ and $\etay$. In the semi-infinite case, details on this classical transformation can be found
in \cite{bergeat1991} and \cite{rutily1992}. In the finite case, the calculation is detailed in \cite{rutily1992}. The results include
the function $g(a,v)$ and the coefficients $R(a,k)$ and $S(a,k)$ as defined by Eqs. (\ref{eq24})-(\ref{eq26}).
One obtains, in a semi-infinite medium
\begin{multline}
\label{eq69}
\eta(a,\tau,z) =\frac{k R(a,k)}{H(a,+1/k)} \frac{\exp(-k\tau)}{1-kz} + \mathrm{Y} [\Re(z)] H(a,-z) \frac{\exp(-\tau/z)}{z} \\
+ \frac{a}{2} \int_{0}^{1} (g/H) (a,v) \exp(-\tau/v) \frac{\d v}{v-z}
\end{multline}
for $z \in \mathbb{C} \setminus \{[0,1] \cup \{+1/k\} \}$,
\begin{equation}
\label{eq70}
\eta(a,\tau,+1) = \frac{k R(a,k)}{H(a,+1/k)} \frac{\exp(-k\tau)}{1-k} + \frac{a}{2} \int_{0}^{1} (g/H)(a,v) \exp(-\tau/v) \frac{\d v}{v-1}
\end{equation}
at $z = +1$,
\begin{multline}
\label{eq71}
\eta(a,\tau,u) = \frac{k R(a,k)}{H(a,+1/k)} \frac{\exp(-k\tau)}{1-ku} + (gT/H)(a,u) \frac{\exp(-\tau/u)}{u} \\
+ \frac{a}{2} \CauchyIntX (g/H)(a,v) \exp(-\tau/v) \frac{\d v}{v-u}
\end{multline}
for $z = u \in ]0, 1[$, and
\begin{multline}
\label{eq72}
\eta(a,\tau,+1/k) = - \frac{k R(a,k)}{H(a,+1/k)}  \left[ S(a,k) - k\tau + \frac{1}{k} \frac{H\p(a,+1/k)}{H(a,+1/k)} \right] \exp(-k\tau) \\
- \frac{a}{2} k \int_{0}^{1} (g/H)(a,v) \exp(-\tau/v) \frac{\d v}{1-kv}
\end{multline}
at $z = +1/k$.\\
 In a finite medium, we need only the expressions of the $\etax$-function, since  
\begin{equation}
\label{eq73}
\etay (a,b,\tau,z) = \etax (a,b,b-\tau,-z)
\end{equation}
due to Eqs. (\ref{eq67})-(\ref{eq68}) and (\ref{eq11}). These are:
\begin{multline}
\label{eq74}
\etax (a,b,\tau,z) = \\
k R(a,k) \left\{ [1 - \ksix (a,b,+1/k)] \frac{\exp(-k\tau)}{1-kz} - \ksiy (a,b,+1/k) \frac{\exp[-k(b-\tau)]}{1+kz} \right\} \\
+ \frac{1}{z} \frac{1}{T(a,z)} \left\{ \mathrm{Y}[\Re(z)] [1 - \ksix (a,b,z)] \exp(-\tau/z) \right. \\
\left. + \mathrm{Y}[-\Re(z)] \ksiy (a,b,-z) \exp[(b -\tau)/z] \right\} \\
+ \frac{a}{2} \int_{0}^{1} g(a,v) \left\{ [1 - \ksix (a,b,v)] \frac{\exp(-\tau/v)}{v-z} - \ksiy (a,b,v) \frac{\exp[-(b-\tau)/v]}{v+z} \right\} \d v
\end{multline}
for $z \in \mathbb{C} \setminus \{[-1,+1] \cup \{\pm 1/k\} \}$,
\begin{multline}
\label{eq75}
\etax (a,b,\tau,-1) = \\
k R(a,k) \left\{ [1 - \ksix (a,b,+1/k)] \frac{\exp(-k\tau)}{1+k} - \ksiy (a,b,+1/k) \frac{\exp[-k(b-\tau)]}{1-k} \right\} \\
+ \frac{a}{2} \int_{0}^{1} g(a,v) \left\{ [1 - \ksix (a,b,v)] \frac{\exp(-\tau/v)}{1+v} + \xi _{\mathrm{Y}} (a,b,v) \frac{\exp[-(b-\tau)/v]}{1-v} \right\} \! \d v ,
\end{multline}
\begin{multline}
\label{eq76}
\etax (a,b,\tau,+1) = \\
k R(a,k) \left\{ [1 - \ksix (a,b,+1/k)] \frac{\exp(-k\tau)}{1-k} - \ksiy (a,b,+1/k) \frac{\exp[-k(b-\tau)]}{1+k} \right\} \\
- \frac{a}{2} \int_{0}^{1} g(a,v) \left\{ [1 - \ksix (a,b,v)] \frac{\exp(-\tau/v)}{1-v} + \ksiy (a,b,v) \frac{\exp[-(b-\tau)/v]}{1+v} \right\} \d v
\end{multline}
at $z = \pm 1$,
\begin{multline}
\label{eq77}
\etax (a,b,\tau,u) = \\
k R(a,k) \left\{ [1 - \ksix (a,b,+1/k)] \frac{\exp(-k\tau)}{1-ku} - \ksiy (a,b,+1/k) \frac{\exp[-k(b-\tau)]}{1+ku} \right\} \\
+ \frac{1}{u} (gT)(a,u) \left\{ \mathrm{Y}(u) [1 - \ksix (a,b,u)] \exp(-\tau/u) \right. \\
\left. + \mathrm{Y}(-u) \ksiy (a,b,-u) \exp[(b-\tau)/u] \right\} \\
+ \frac{a}{2} \CauchyIntX g(a,v) \left\{ [1 - \ksix (a,b,v)] \frac{\exp(-\tau/v)}{v-u} - \ksiy (a,b,v) \frac{\exp[-(b-\tau)/v]}{v+u} \right\} \d v ,
\end{multline}
for $z = u \in ]-1,0[ \cup ]0,+1[$, and
\begin{multline}
\label{eq78}
\etax (a,b,\tau,-1/k) = k R(a,k) \left\{ \frac{1}{2} [1 - \ksix (a,b,+1/k)] \exp(-k\tau) \right. \\
+ [S(a,k) - k(b-\tau)] \ksiy (a,b,+1/k) \exp[-k(b-\tau)] \\
\left. - \frac{1}{k} \ksiy\p (a,b,+1/k) \exp[-k(b-\tau)] \right\} \\
+ \frac{a}{2} k \! \int_{0}^{1} \! g(a,v) \! \left\{ [1 - \ksix (a,b,v)] \frac{\exp(-\tau/v)}{1+kv} + \ksiy (a,b,v) \frac{\exp[-(b-\tau)/v]}{1-kv} \right\} \! \d v ,
\end{multline}
\begin{multline}
\label{eq79}
\etax (a,b,\tau,+1/k) = - k R(a,k) \left\{ [S(a,k) - k\tau] [1 - \ksix (a,b,+1/k)] \exp(-k\tau) \phantom{\frac{1}{2}} \right. \\
\left. + \frac{1}{k} \ksix\p (a,b,+1/k) \exp(-k\tau) + \frac{1}{2} \ksiy (a,b,+1/k) \exp[-k(b-\tau)] \right\} \\
- \frac{a}{2} k \! \int_{0}^{1} \! g(a,v) \! \left\{ [1 - \ksix (a,b,v)] \frac{\exp(-\tau/v)}{1-kv} + \ksiy (a,b,v) \frac{\exp[-(b-\tau)/v]}{1+kv} \right\} \! \d v
\end{multline}
at $z = \pm 1/k$. It is worth noting that these expressions only involve the functions $\ksix$ and $\ksiy$, not the functions $H$, $X$ or $Y$. They can be checked putting in $b \to + \infty$, since then $\etax$ does coincide with $\eta$ and $\etay$ vanishes. \\
 From the physical interpretation of the functions $\eta$, $\etax$ and $\etay$ we reminded in
Sec. 5 of \cite{rutily1994}, it follows that these functions are continuous on the right at $\tau = 0$, and
that $\etax$ and $\etay$ are continuous on the left at $\tau = b$. Their restrictions to the boundary
layers are thus given by the limits of the above expressions as $\tau \to 0$ and $\tau \to b$, which
coincide with their values at $\tau = 0$ and $\tau = b$. Using the integral equations of Secs. \ref{sec3}
and \ref{sec4}, one obtains the following surface expressions of the functions $\eta$, $\etax$ and $\etay$,
valid for any $z$ in $\mathbb{C}^*$: in a semi-infinite medium
\begin{equation}
\label{eq80}
\eta(a,0,z) = \frac{1}{z} \left\{ 1 - \mathrm{Y}[-\Re(z)] H(a,-z) \right\} ,
\end{equation}
and in a finite medium
\begin{equation}
\label{eq81}
\etax (a,b,0,z) = \frac{1}{z} \left\{ 1 - \mathrm{Y}[-\Re(z)] X(a,b,-z) \right\} ,
\end{equation}
\begin{equation}
\label{eq82}
\etax (a,b,b,z) = \frac{1}{z} \mathrm{Y} [\Re(z)] Y(a,b,z) , 
\end{equation}
so that, from Eq. (\ref{eq73})
\begin{equation}
\label{eq83}
\etay (a,b,0,z) = - \frac{1}{z} \mathrm{Y}[-\Re(z)] Y(a,b,-z) ,
\end{equation}
\begin{equation}
\label{eq84}
\etay (a,b,b,z) = - \frac{1}{z} \left\{ 1 - \mathrm{Y}[\Re(z)] X(a,b,z) \right\} .
\end{equation}
 These expressions still hold on the imaginary axis, since $\mathrm{Y}(0) = 1/2$. Substituting
them into the relations (\ref{eq64}) and (\ref{eq66}) at $\tau = 0$ and $\tau = b$, one obtains the following
surface values of the $B$-function:
\begin{equation}
\label{eq85}
B(a,0,z) = H(a,z) \qquad [\Re(1/z) > - k] ,
\end{equation}
\begin{equation}
\label{eq86}
B(a,b,0,z) = X(a,b,z) \qquad (z \in \mathbb{C}^*) ,
\end{equation}
\begin{equation}
\label{eq87}
B(a,b,b,z) = Y(a,b,z) \qquad (z \in \mathbb{C}^*) ,
\end{equation}
which generalize in the complex plane the classical relations (\ref{eq62})-(\ref{eq63}) derived over $]0,1]$
 more than half a century ago. The need for such an extension is explained at the end
of Sec. 2.3 of \cite{bergeat1998}. We conclude that the integral equations of Secs. \ref{sec3} and \ref{sec4} provide the
connection between the internal and the surface solution to the albedo problem (\ref{eq61}) with complex parameter $z$.\\
 In conclusion, we outline the bibliography on the exact solution of the albedo problem. We have seen that the surface solution (\ref{eq62}) was given for the first time by Ambartsumian in a half-space \cite{ambartsumian1942}. The internal solution was first derived in the context of neutron transport by means of Case's singular eigenfunction method \cite{kuscer1964,case1967}. It was
expressed in terms of the $\eta$-function by Danielian and Mnatsakanian \cite{danielan1975}. The functions $F$ and $\tilde{F}$ of these authors are indeed related to our function $\eta$ by $F(a,\tau,u) = u \eta(a,\tau,u)$ and $\tilde{F}(a,\tau,u) = u \eta(a,\tau,-u)$ ($0 < u \leq +1$). Similar functions have also been introduced by Crosbie and Linsenbardt \cite{crosbie1977} and Viik \cite{viik1984}. The first analytical expression of the $\eta$-function is in Danielian and Pikichian \cite{danielan1977}, whose relation (36) is valid over $[-1,0[$, not over $]0,1]$.\\
 In the finite case, the surface solution (\ref{eq63}) was given by Busbridge \cite{busbridge1955b}, and
then Case's method was again required for calculting the internal field \cite{mccormick1964}. The first
expression in terms of the $\etax$- and $\etay$-functions is due to Kagiwada and Kalaba \cite{kagiwada1968,kagiwada1975},
who used the invariant imbedding method. Their functions $b$ and $h$ coincide with our
functions $\etay$ and $\etax$ respectively. Danielian's approach in \cite{danielan1976} is similar, since his
functions $F$ and $\tilde{F}$ are related to $\etax$ and $\etay$ by $F(a,b,\tau,u) = u \etax(a,b,\tau,u)$ 
and $\tilde{F}(a,b,\tau,u) = u \etax(a,b,\tau,-u) = u \etay(a,b,b-\tau,u)$ ($0 < u \leq 1$). In \cite{danielan1983}, Danielian wrote the
analytical expressions of his functions $F$ and $\tilde{F}$, henceforth denoted by $f$ and $\tilde{f}$ in a finite slab.
His formulae may be simplified with the help of Eqs. (\ref{eq40})-(\ref{eq41}), since they contain two times too many terms.
We recall that Danielian was not aware of Eqs. (\ref{eq40}) and (\ref{eq41}) {\em separately\/} (see the end of Sec. \ref{sec4}).
After simplification, Eqs. (11)-(12) of \cite{danielan1983} do yield our own Eq. (\ref{eq77}) again.
The latter and accompanying equations in the series (\ref{eq74})-(\ref{eq79}) are given in \cite{rutily1992}.

\section{Conclusion}\label{sec6}

 The integral equations satisfied by the $H$-function (Sec. \ref{sec3}) and by the $X$- and $Y$-functions (Sec. \ref{sec4}) are our main results.
The former one generalizes, in the whole complex plane, an integral equation derived by Busbridge \cite{busbridge1955a} over $[-1,+1]$.
The latter ones for the $X$- and $Y$-functions generalize and simplify two integral equations derived
by Danielian \cite{danielan1983}. These integral equations clarify the slab albedo problem (\ref{eq61}), since
they show that its internal solution simplifies on the boundary layers into the surface
solution.\\ 
 Contrary to Danielian \cite{danielan1983}, we consider that these integral equations are of little interest for computing the functions $H$, $X$, $Y$, $\ksix$ and $\ksiy$. The $H$-function can be easily evaluated from one of its many analytical expressions. This approach is particularly interesting in strongly scattering media. The best way we know to compute the functions
$X$, $Y$, $\ksix$ and $\ksiy$ is that developed by Busbridge \cite{busbridge1955a} and Mullikin et al. \cite{mullikin1964,carlsted1966}.
The four functions are derived from some simpler auxiliary functions that solve Fredholm integral equations with regular kernels.
This approach is that of Danielian himself in two important papers \cite{danielan1993,danielan1994} he published after Ref. \cite{danielan1983}. It was also
developed in Ref. \cite{rutily1992}, which contains ten-figure tables of the functions in question.

\end{document}